\documentclass[journal]{vgtc}         \ifpdf
  \pdfoutput=1\relax                   
  \usepackage{graphicx}                
  \DeclareGraphicsExtensions{.pdf,.png,.jpg,.jpeg} 
\else
  \usepackage{graphicx}                
  \DeclareGraphicsExtensions{.eps}     
\fi%

\graphicspath{{figures/}{pictures/}{images/}{./}} 

\usepackage{microtype}                 
\PassOptionsToPackage{warn}{textcomp}  
\usepackage{textcomp}                  
\usepackage{mathptmx}                  
\usepackage{times}                     
\usepackage{cite}                      
\usepackage{tabu}                      
\usepackage{booktabs}                  
\usepackage{verbatim}
\usepackage{hyperref}

\usepackage{amsmath,array}
\usepackage{amssymb}
\usepackage{enumitem}

\usepackage[version=4]{mhchem}

\def\lhs{\mathcal{L}}
\def\rhs{\mathcal{R}}

\makeatletter
\newsavebox\myboxA
\newsavebox\myboxB
\newlength\mylenA

\newcommand*\xoverline[2][0.75]{%
    \sbox{\myboxA}{$\m@th#2$}%
    \setbox\myboxB\null
    \ht\myboxB=\ht\myboxA%
    \dp\myboxB=\dp\myboxA%
    \wd\myboxB=#1\wd\myboxA
    \sbox\myboxB{$\m@th\overline{\copy\myboxB}$}
    \setlength\mylenA{\the\wd\myboxA}
    \addtolength\mylenA{-\the\wd\myboxB}%
    \ifdim\wd\myboxB<\wd\myboxA%
       \rlap{\hskip 0.5\mylenA\usebox\myboxB}{\usebox\myboxA}%
    \else
        \hskip -0.5\mylenA\rlap{\usebox\myboxA}{\hskip 0.5\mylenA\usebox\myboxB}%
    \fi}
\makeatother

\onlineid{0}

\vgtccategory{Research}
\vgtcpapertype{application/design study}

\title{Dynamic Influence Networks for Rule-based Models}

\author{Angus G. Forbes, Andrew Burks, Kristine Lee, Xing Li, Pierre Boutillier, Jean Krivine, and Walter Fontana}
\authorfooter{
\item Angus G. Forbes is with University of California, Santa Cruz. E-mail: angus@ucsc.edu. Website: \url{https://creativecoding.soe.ucsc.edu/}
\item Andrew Burks, Kristine Lee, and Xing Li are with University of Illinois at Chicago. E-mail: \{aburks3, khlee2, xli227\}@uic.edu. 
\item Jean Krivine is with Universit\'e Paris Diderot. E-mail: jean.krivine@irif.fr. 
\item Pierre Boutillier and Walter Fontana are with Harvard Medical School. E-mail: \{pierre\_boutillier, walter\}@hms.harvard.edu. 
}

\shortauthortitle{Forbes \MakeLowercase{\textit{et al.}}: Dynamic Influence Networks}

\abstract{We introduce the Dynamic Influence Network (DIN), a novel visual analytics technique for representing and analyzing rule-based models of protein-protein interaction networks. Rule-based modeling has proved instrumental in developing biological models that are concise, comprehensible, easily extensible, and that mitigate the combinatorial complexity of multi-state and multi-component biological molecules. Our technique visualizes the dynamics of these rules as they evolve over time. Using the data produced by \textit{KaSim}, an open source stochastic simulator of rule-based models written in the \textit{Kappa} language, DINs provide a node-link diagram that represents the influence that each rule has on the other rules. That is, rather than representing individual biological components or types, we instead represent the rules about them (as nodes) and the current influence of these rules (as links). Using our interactive \textit{DIN-Viz} software tool, researchers are able to query this dynamic network to find meaningful patterns about biological processes, and to identify salient aspects of complex rule-based models. To evaluate the effectiveness of our approach, we investigate a simulation of a circadian clock model that illustrates the oscillatory behavior of the KaiC protein phosphorylation cycle.
} 


\CCScatlist{ 
 \CCScat{K.6.1}{Management of Computing and Information Systems}%
{Project and People Management}{Life Cycle};
 \CCScat{K.7.m}{The Computing Profession}{Miscellaneous}{Ethics}
}

\teaser{
  \centering
  \includegraphics[width=6.56in]{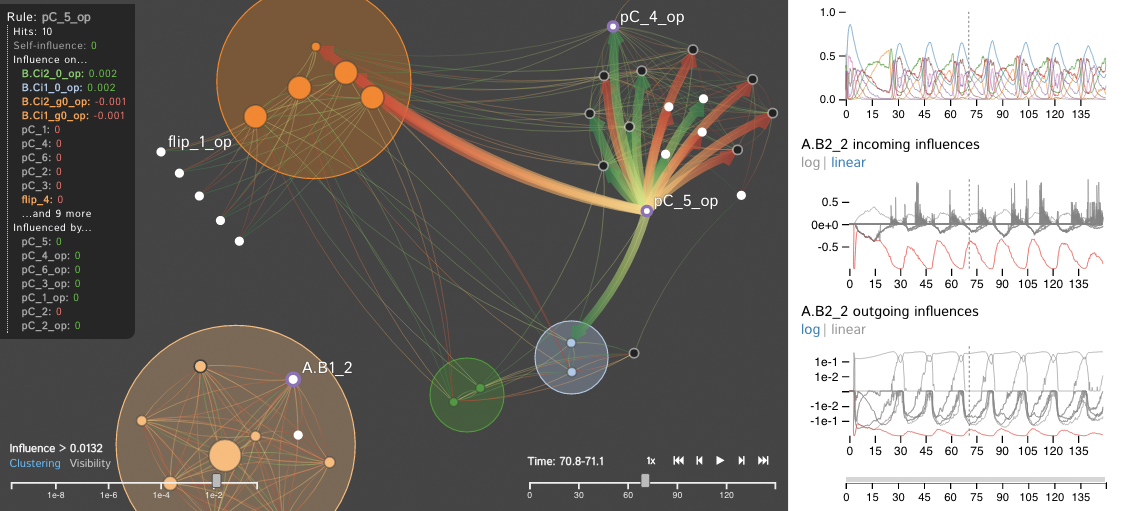}
  \caption{A screenshot of the \textit{DIN-Viz} application for analyzing the dynamics of a rule-based model of a protein-protein interaction network (i.e., a Dynamic Influence Network). Our approach emphasizes the influence rules have on each other and enables users to analyze the dynamics of these influences as they change over time. The left panel shows a network of interconnected rules at a specific time step. The right panel provides a global overview of the system as well as detailed information about selected rules.}
	\label{fig:teaser}
}

\vgtcinsertpkg

\begin{document}

\firstsection{Introduction}

\maketitle


\noindent In an oft-cited overview of the goals of computational systems biology, Hiroaki Kitano explains the role of simulation in visualizing and understanding complex systems. Simulation, he writes, can ``predict the dynamics of systems so that the validity of the underlying assumptions can be tested''~\cite{kitano2002computational}. Furthermore, he articulates the importance of dynamic diagrams that help to identify patterns in biological systems, to explain why they emerge, and to enable researchers to understand how they can be controlled~\cite{kitano2002systems}. In this paper, we introduce the Dynamic Influence Network, a novel representation that enables the effective visualization of the dynamics of a system of rules to describe complex biological processes. 

A central challenge across complex systems is to understand how a multitude of heterogenous and interacting agents can give rise to coherent system behavior. The nature of the challenge is as much empirical as it is theoretical. It is empirical, because the relevant agents of a system, their interactions, and the global behavior they induce need to be discovered and characterized. It is theoretical, because the connections between agent-centric interactions and system-level behavior need to be grasped at a level beyond mere description, so as to separate contingency from necessity.

The technical challenge in modeling, even at a high level of abstraction, a complex system like a cell is that the molecular components and their manifold interactions often result in an overwhelming combinatorial complexity, frustrating an initial insight needed for guiding the construction of a model. The challenge is particularly acute for dynamical models of a mechanistic kind, which are preferred over statistical black boxes when a causal account of system behavior is sought in terms of interactions between molecular agents. The molecular level is privileged because of the fundamental role it plays in our understanding of cellular processes during development, disease, and evolution.


A specific case in point are molecular signaling processes in which criss-crossing cascades of interactions between proteins are triggered in response to exogenous or endogenous molecular events. These cascades integrate a multitude of signals through which a cell homes in on an appropriate adaptive response, such as moving, dividing, differentiating, repairing, maintaining homeostasis, and possibly attacking itself or other cells.

Proteins are large molecules that mediate these signaling cascades by virtue of chemical tags attached to specific locations (sites) at their surface. These tags influence a protein's behavior. For example, a phosphorylation can change the conformation of a protein, thereby modulating its chemical activity or altering which other proteins it can transiently recognize and associate with. The set of tags, i.e.\@ the state of a protein, is in turn modified through interaction with other proteins whose behavior is controlled by their own tagging state.

\label{ss:Rule-based Modeling}
\subsection{The \textit{Kappa} Language for Rule-based Modeling}

Information about the state conditions required for an interaction between specific proteins can be extracted or inferred from a rapidly growing experimental literature and curated databases. The \textit{Kappa} language~\cite{danos2004formal} is designed to formally express such interactions as rules that can be executed by machine. The approach follows the representational scheme of chemistry where graphs are used to express molecules as connected atoms and graph-rewrite rules capture local bond rearrangements that result in chemical reactions. There is a crucial distinction between a \textit{mechanism}, in which parts of entities are responsible for a specific transformation, and a \textit{reaction}, in which molecular species (the full entities) affected by the transformation are produced or consumed. A \textit{Kappa} rule is a schema that does not require a full specification of the interacting entities, referring only to the parts that are necessary for the transformation. At the heart of rule-based modeling is the \emph{agent abstraction}, which conceptualizes a protein as an agent defined by a name and an interface of sites that represent distinct interaction capabilities, e.g.\@ binding and modification, without explicitly representing the underlying structural or chemical features that enable these capabilities~\cite{danos2004formal,chylek2014rule,chylek2015modeling}. 

\textit{KaSim} is a stochastic simulator of rule-based models written in the \textit{Kappa} language~\cite{KaSimKaSaManual2017}. \textit{Kappa} expresses only local rules of action, which are both ``descriptions of mechanistic knowledge and executable instructions''~\cite{feret2009internal}.
Since the definitions are completely local and not based on any kind
of enumeration, the simulation algorithm that executes these rules has a per event time cost which is independent
of the size of the set of generable species, or element types, and independent of the total size of the elements or agents in the system. Even on a standard desktop computer, \textit{KaSim} can easily generate simulations of a system with millions of agents and that is defined by hundreds of rules~\cite{danos2007scalable}. The output of a \textit{KaSim} simulation, which is ingested into our \textit{DIN-Viz} software, contains information about: a) a series of \textit{observables}, the population of elements or groups of elements within the system; as well as b) the number of times individual rules fire over time, or alternately, the probability that they fire in different configurations. We explicitly define this \textit{influence} metric for rule interactions in \autoref{ss:Define}.

\subsection{Visual Analysis for Understanding Rule-based Models}
\label{ss:VisualModeling}
The rule-based approach to models proceeds without a preconceived notion of how the global behavior of interest arises in a system. In traditional models, e.g., reaction-based models, one often has a mechanism in mind and aims at proving that it indeed behaves as claimed. If such initial insight is not available, one may proceed by carefully expressing known facts about interactions between parts of entities. The result is a collection of local rules. The problem is that \emph{even if the sought-after behavior is produced, we still need to work out why it behaves that way.} That is, a rule-based approach often ends up replacing a world we don't understand with a model we don't understand. In this regard, the modeler is in the same position as an experimentalist, but with the much easier task of dealing with a system about which everything at the micro-level is known, allowing one to focus on how the micro-level gives rise to the observed macro behavior. An advantage of rule-based models is the ease by which the assumptions underlying such models can be altered and the consequences explored.

Rule-based models are ideally suited for mechanistic causal analysis, that is, for identifying the chains of events (rule applications) that were necessary in obtaining a particular result. The formalization of a biologically meaningful notion of mechanistic causality that permits extraction of pathways, or workflows, from a simulation of a rule-based model is a challenging subject of ongoing research~\cite{cohen2015darpa}. Systems biologists utilize simulation for a range of applications, but broadly speaking, high-level goals for using rule-based modeling to simulate complex systems include the following goals, related to accessibility, execution, and understanding~\cite{chylek2015modeling, danos2009rule, krakauer2011challenges}: 

\begin{enumerate}[noitemsep,leftmargin=*]
\item [\textbf{G1}] Understand how a multitude of heterogenous and interacting agents can give rise to coherent system behavior; 
\item [\textbf{G2}] Utilize dynamical models for causal explanations of biochemical interactions;
\item [\textbf{G3}] Provide effective visual analytics methods for investigating simulation outputs of rule-based models.  
\end{enumerate}

Protein-protein interaction networks with hundreds to hundreds of thousands of individual elements can be difficult to represent. However, visualizing the rules that govern these elements and their interactions with each other or with other elements provides a more concise way to represent the system. This highlights the salient aspects of the system, while at the same time accurately aggregating the lower-level elements. 

Through our survey of the relevant literature in systems biology and information visualization (see \autoref{RelatedWorkSection}), along with in depth conversations with domain experts in rule-based modeling, we articulate the main visual analytics tasks that an interactive visualization of dynamic rule influences should support:

\begin{enumerate}[noitemsep,leftmargin=*]
\item [\textbf{T1}] Identify influence patterns between rules and within groups of rules;
\item [\textbf{T2}] Visualize the dynamics of rules as they change over time;
\item [\textbf{T3}] Generate hypotheses about the relationship between local interactions and their participation in global behaviors; 
\item [\textbf{T4}] Annotate components of the dynamic system in order to highlight salient features and share them with others;
\end{enumerate}

We introduce a new visual analytics technique, the Dynamic Influence Network, or DIN for short, to aid in the understanding of how coherent system behavior emerges from a set of rules. Additionally, we present a software tool, \textit{DIN-Viz}, that provides systems biologists with the means to visually examine dynamic data collected during simulation.\footnote{\textit{DIN-Viz} is open source software freely available at \url{https://github.com/CreativeCodingLab/DynamicInfluenceNetworks}.} The paper thus presents the following contributions:

\begin{itemize}[noitemsep,leftmargin=*]
\item{We introduce a dynamic network representation of rules and rule influence (\autoref{ss:Define});} 
\item{We provide a web-based software application, created in JavaScript using the D3.js library~\cite{bostock2011d3}, for interactively exploring and analyzing Dynamic Influence Networks (\autoref{dinviz});}
\item{We present a thorough explication of a use case that clearly articulates the various strategies a systems biologist uses to make sense of a rule-based model for a complex dynamic system (\autoref{UseCaseSection}).}
\end{itemize}

\noindent Additionally, \autoref{RelatedWorkSection} discusses relevant related work from biological modeling, biological data visualization, causality visualization, and dynamic graph visualization, and \autoref{DiscussionSection} presents user feedback from domain experts in systems biology and molecular biology.

\section{Related Work} \label{RelatedWorkSection}

Visual analytics often involves mitigating the complexity of datasets that describe systems that are too large or too complicated to reason about without external representation. This is true as well for visualization projects that aim to explore biological pathways. Instead of representing individual proteins or biological elements, which could number in the tens or hundreds of thousands for a single cell, or even the types of elements, which could also require a large number of nodes, our approach visualizes the relevant \textit{rules} that govern the interactions between elements. Since most systems have many less rules than elements or element types, this decreases the amount of information that needs to be represented (although issues related to uncertainty, incompleteness, and complexity can still occur with rule-based approaches). Moreover, this information is potentially more useful for gaining insight into how a system works, and for validating useful models that can guide research into biological pathways.

\subsection{Biological Modeling}

While the majority of pathway visualization tools apply to reaction-based modeling, our project introduces a visual analysis technique for rule-based models. In addition to the \textit{Kappa} language described above, the \textit{BioNetGen} language~\cite{faeder2009rule,harris2016bionetgen} is also widely used by systems biologists. While the differences between these languages are minimal, one benefit of \textit{Kappa}, according to Wilson et al.~\cite{wilson2015kappa}, is that tools in the \textit{Kappa} ecosystem make use of formal methods to aid in information discovery and in debugging models.

A modeling framework called \textit{PySB} aims to make it easier to build mathematical models of biochemical systems as Python programs~\cite{lopez2013programming}. In their approach, models are not only created using programs, these models are already executable programs. \textit{PySB} transforms the Python code into either \textit{BioNetGen} or \textit{Kappa} rules, and provides methods that make it easier to create macros that encode recurrent biochemical patterns and to define complex networks as reusable modules. Pedersen~et~al.~\cite{pedersen2015high} also introduce a modular extension to \textit{Kappa} that provides a means for writing modular rule-based models.

While our technique was created using \textit{Kappa} rules and \textit{KaSim} outputs, it should be straightforward to extend it to other rule-based modeling languages or high-level modeling frameworks.

\subsection{Dynamic Network Visualization}

Efforts to effectively visualize graphs with nodes or edges that represent temporal data or that have a topology that evolves over time are cataloged by Beck et al.~\cite{beck2017taxonomyCGF}. They survey the landscape of dynamic graph visualization, categorizing projects primarily in terms of how they represent time, that is, whether or not they use animation or a static timeline to show the evolution of networks. These categories are then further parcellated according to which layout strategies they utilize and how they address particular problems inherent in dynamic datasets. Inspired by Moody et al.~\cite{moody2005dynamic}, who investigate animated network ``movies'' for a range of sociological datasets, our tool features an animated node-link diagram whose layout is determined by clusters of influence (measured by how likely rules are to fire at the same time), either on a per-frame basis or within a user-selected time window. An interactive timeline is used to navigate through time, and more detailed information about selected nodes is presented for the currently selected time period.

Interesting recent approaches to visualizing dynamic data include Archambault and Purchase's work on dynamic attribute cascades~\cite{Archambault2016_dynamicAttributeCascades}, Mashima et al.'s \textit{GMap}~\cite{mashima2012visualizing}, and Bach et al.'s \textit{GraphDiaries}~\cite{bach2014graphdiaries} and \textit{Matrix Cubes}~\cite{bach2014visualizing} techniques. However, since a main goal of our visualization was to emphasize the relationship of rules to other rules, we elected to use a visual representation that made it easier to apply visual encodings to the links between nodes (see \autoref{DIN_Section}). Techniques by Ma et al.~\cite{Ma_EuroVis_SP,Ma2015_JIST_SwordPlots}, Purgato et al.~\cite{Purgato2017_BHI}, and Ye et al.~\cite{YeNeuroImage2015} present multiple synchronized representations of a dynamic brain network to provide additional insight into the community dynamics within the network. Our tool also presents auxiliary representations to support the analysis of dynamic data, providing detail on demand for selected nodes.

Vehlow et al.~\cite{vehlow2016visualizing} provide a thorough overview of different approaches to grouping data within graphs. A taxonomy of methods categorizes groups as juxtaposed, embedded, superimposed, or encoded using visual node attributes. Our technique utilizes superimposition, providing colored clusters as a way to show group membership of nodes that are similar, as well as visual node attributes, enabling a user to apply coloring to indicate a secondary grouping of nodes. Hadlak et al.~\cite{hadlak2013supporting} present a network layout in which each node represents a cluster, and contains a time plot providing an overview of the temporal trend of the cluster, as well as a secondary view that shows time series data describing changes to a selected cluster. Our software tool also allows the user to examine more detailed information about temporal trends within the network.

\subsection{Causality Visualization}

Elmqvist and Tsigas~\cite{Elmqvist2004a} introduce the Growing Squares and Growing Polygons techniques to explore causality, finding that they are significantly more useful than static graphs or Hesse diagrams for reasoning about systems. In these techniques, differently sized shapes are used to indicate information flow in a system of interacting processes, filling with different colors to represent the changing influence of the different processes. 

Ware et al.\cite{ware1999visualizing} explore the use of visual causal vectors to indicate causal relationships between data elements, and Bartram and Yao~\cite{bartram2008animating} utilize animated causal overlays in order to highlight causal flows and to indicate the relative strength of the causal effect. Kadaba et al.~\cite{kadaba2007visualizing} also find that the use of animation is superior in terms of both accuracy and speed in comparison to static representations for facilitating comprehension of complex causal relations.  

More recently, Dang et al.~\cite{biovisDang2} introduce \textit{ReactionFlow} to highlight the inherent causality in biological pathways. In this technique, different reactions are presented in a column at the center of the screen, and a user can select any reaction, or any protein involved in a reaction, and play an animation that indicates which other reactions could occur downstream. Zhang et al.~\cite{zhang2015visualizing} introduce the \textit{ReView} tool for visualizing traffic causality and to reason about the origin of network traffic anomalies that may be indicative of suspicious requests. Also intended to help determine root causes of events (although in an entirely different domain), Vigueras and Botia \cite{vigueras2007tracking} introduce an algorithm for defining causality graphs for debugging multi-agent systems in order to track the causality of events produced by interactions among agents in a group. Forbes et al.~\cite{forbes2010behaviorism} also investigate the use of an animation framework to represent simulations of dynamic systems. Our technique aims to provide systems biologists with the ability to more effectively reason about the causal mechanisms within protein-protein interaction networks.

\subsection{Biological Network Visualization}

Murray et al.~\cite{MurrayBMC2017} survey biological pathway visualization projects and introduce a taxonomy of visualization tasks for the analysis of biological pathway data. Relevant tasks are organized in a high-level categorization as \textit{attribute} tasks, \textit{relationship} tasks, and \textit{modification} tasks. Interestingly, simulations of rule-based models are not explicitly discussed. However, this taxonomy does describe tasks that are relevant to visualizations of rule-based models, and our approach in particular, such as \textit{grouping}~\cite{biovisPaduano1,Paduano2016_VINCI,Vehlow2015BMCBIO}, \textit{annotation}~\cite{Dang2017_BioLinker_IEEEPacificVis,Dang2016a_IEEEEuroVis,dinkla2014examine,lex2013entourage,funahashi2008celldesigner}, and \textit{causality}~\cite{Partl2013enroute,felciano2013predictive,Kramer2013ipacausal}.

A range of projects introduce techniques for visualizing rule-based models of protein-protein interaction networks. Danos et al.~\cite{danos2007rule} discuss the use of \textit{contact maps} to visually represent stochastic trajectories for user-defined observables in order to tell a ``story'' that summarizes how a given event type can be obtained. Kohn et al.~\cite{kohn2006depicting} introduce \textit{molecular interaction maps} that explicitly define the topology of rule-based networks, and that can be used for simulating the interaction of molecular rules. Inspired by these efforts, as well as by the earlier process diagrams of Kitano et al.~\cite{kitano2005using}, Chylek et al.~\cite{chylek2011guidelines} provide guidelines for visualizing and annotating rule-based models using interactive \textit{extended contact maps} that represent a cell signaling system so that it is both visual and executable.

Smith et al.'s \textit{RuleBender}~\cite{smith2012rulebender} provides a framework for editing and exploring rule-based systems of intracellular biochemistry, such as \textit{BioNetGen}. In this framework, the primary visualization provides an interactive contact map in which  molecules are rendered as large gray nodes and domain states are positioned inside these nodes. Rules are represented as links between specific sites within the nodes, and nested isocontours are used to define a compartment hierarchy of elements related to particular components of the cell. A secondary visualization shows the relations between the reaction rules that describe the behavior of a system. That is, similar to our approach, the influence graph shows if a rule activates or inhibits another rule. However, our approach further emphasizes how these influences can change over time, facilitating the analysis of the dynamics of the system. 

Gostner et al.~\cite{gostner2015graphical} provide an overview of graphical modeling software tools for representing or simulating reaction-based models, and detailing usability and perceptual issues that they can introduce. They find that creating graphical languages with many glyphs reduces ambiguity, but at the cost of introducing visual clutter and making the visual layout unappealing to users. They advocate for a minimalist approach that presents only the visual elements necessary for a particular analysis. Our tool also uses a smaller visual language for representing rules and clusters of rules.

Pedersen et al. introduce \textit{Bio Simulators}~\cite{pedersen2014bio}, a web-based framework that uses \textit{KaSim} to run simulations of \textit{Kappa} rules. The visualization output is a simple chart that shows the population of predefined ``observables'' within the system, that is biological agents (e.g., proteins or protein complexes) that are affected by the rules. This visualization provides an overview of the system, but does not indicate specifically which rules are responsible for these changes in observables, nor provide insight into how the activity of rules affects other rules. Our \textit{DIN-Viz} tool provides this type of visual output in a secondary panel, providing an alternative perspective of the system.

\section{Dynamic Influence Networks}\label{DIN_Section}
\vspace{-.1cm}

In this section, we introduce details of the Dynamic Influence Network and our visual analytics software tool, \textit{DIN-Viz}. The \textit{DIN-Viz} interface presents two interconnected view panels, the Network Panel and the Data Panel. The Network Panel provides an overview of the influence the rules have upon each other, and the Data Panel provides additional detail both about selected individual rules, as well as an overview of the entire system. These views can be temporally navigated using an interactive timeline. Interactive thresholding defines which nodes and links are visible, and pop-up option panels enable users to emphasize different characteristics of the network and to toggle between different clustering metrics.

By using \textit{DIN-Viz}, researchers can engage in each of the visual analysis tasks defined in \autoref{ss:VisualModeling}. Specifically, our interface aims to facilitate hypothesis generation and validation. We discuss this in detail in \autoref{UseCaseSection}, where we present a walkthrough of an analysis of a complex biological system. Briefly, a user is provided with an overview of the dynamic system in which rules that are tightly coupled are grouped together in clusters (\textbf{T1}). A user can then create custom clusters of specific rules, arranging them in close spatial proximity in order to evince meaningful behaviors at given times (\textbf{T2}). For example, a user may be interested to learn if particular rules fire together, independently of whether or not they influence each other. A user can then temporally traverse the dynamic system in order to identify particular times where this hypothesis about synchronized rule firing is true or not true (\textbf{T3}). The user can apply annotations (labels and color cues) to identify relevant features of the system in order to present this information to others (\textbf{T4}).

\textit{DIN-Viz} was created through a collaborative design process, in which we received expert feedback from our systems biology collaborators over the course of a six month development period. During this time, many iterations of the application and visual encodings were developed and refined. As the software became more adept at representing rule-based models, new datasets were provided that initially challenged our application, both in terms of the robustness of our implementation and the utility of our design choices, forcing us to rethink some aspects of our approach and to extend others. Through this process, \textit{DIN-Viz} has become increasingly effective at representing large rule-based models in order to facilitate analysis tasks that utilize Dynamic Influence Networks.

\begin{figure}[!t]
\begin{center}
\includegraphics[width=\columnwidth]{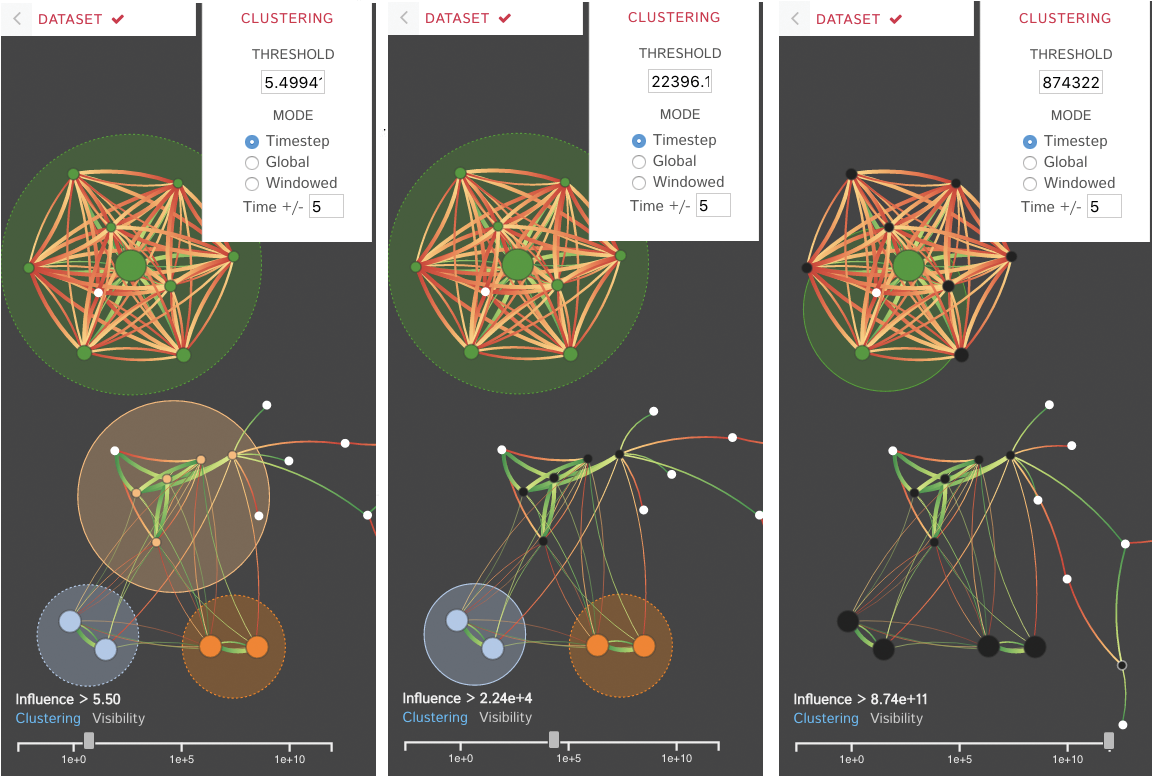}
\caption[ThresholdCluster1]{These screen captures show views of the DIN at the same time step, but each using three different clustering thresholds. As the user increases the clustering threshold value (left to right) using the slider at the bottom-left of the Network Panel, fewer and fewer clusters are created. In the pop-up menu, a user can enter a precise clustering threshold and choose the clustering window options.}
\label{ThresholdCluster1}
\end{center}
\vspace{-0.5cm}
\end{figure}

\subsection{Defining the Dynamic Influence Network} 
\label{ss:Define}

Before describing further details of our visualization technique, we introduce a formal definition of \textit{influence} in relation to \textit{Kappa} rules. 

Let $s$ be a rule
\vspace{-.1cm}
\begin{equation}
s:\quad \lhs_s \longrightarrow \rhs_s \qquad \text{@ }\gamma_s
\end{equation}
with $\lhs_s$ and $\rhs_s$ \textit{Kappa} patterns, and $\gamma_s$ a rate constant. The activity $\alpha_s$ of $s$ indicates the propensity of $s$ to induce the next state transition in a mixture of fully specified molecular species $\mathcal{M}$ (the system). The probability that $s$ induces the next transition is simply the relative propensity, $\alpha_s/\sum_r^N\alpha_r$. Moreover, 
\vspace{-.1cm}
\begin{equation}
\alpha_s=\gamma_s\cdot |\{\lhs_s;\mathcal{M}\}| \cdot\dfrac{1}{\sigma_s} \label{eq:activity}
\end{equation}
where $|\{\lhs_s;\mathcal{M}\}|$ denotes the number of matchings of $\lhs_s$ in the mixture $\mathcal{M}$ and $\sigma_s$ the number of symmetries preserved by $s$. The term $|\{\lhs_s;\mathcal{M}\}|/\sigma_s$ is the number of physical configurations that are distinguishable with respect to the mechanism expressed in $s$. The activity $\alpha_s$ of a rule corresponds to the conventional chemical kinetics notion of flux (velocity) through a reaction. 

The Dynamic Influence Network (DIN) is a directed graph whose nodes are rules. A directed edge from rule $r$ to rule $s$ has a weight that reports the contribution of events due to rule $r$ ($r$-events) on the activity of rule $s$ over some fixed time interval $[t,t+\tau]$. Rule $r$ affects the activity of $s$ by either generating or destroying matchings of $\lhs_s$ in $\mathcal{M}$. Note that in a system of reactions (rather than rules), any firing of the reaction $\ce{A -> B}$ has an immediate impact on the activity of the reaction $\ce{B + C -> D}$, if the system already contains nonzero amounts of $\ce{C}$. In a rule-based system, however, a firing of rule $r: \ce{A(x_u) -> A(x_p)}$ need not have an immediate impact on the activity of $s: \ce{A(x_p,y_p) -> A(x_p,y_u)}$. When $r$ fires on an instance $\ce{A(x_u,y_u)}$ in the mixture $\mathcal{M}$, it has no immediate impact on the activity of $s$; it does, however, when it fires on $\ce{A(x_u,y_p)}$. 

Let $i$ index the sequence of events that occur in $[t,t+\tau]$. The $i$th event causes a transition of the system from state $M_i$ (before the event) to state $M_{i+1}$ (after the event). If the $i$th event is due to an application of $r$, the contribution of $r$ to the \emph{relative change in activity} of rule $s$ is given by
\vspace{-.1cm} 
\begin{eqnarray}
\Delta_{i}(r\rightsquigarrow s)=\begin{cases}
\dfrac{\alpha_s(i+1)-\alpha_s(i)}{\alpha_s(i)} &\text{ if $i$ is due to $r$ and } \alpha_s(i)\ne 0\\
0 &\text{ otherwise}
\end{cases}\label{eq:relative}
\end{eqnarray}
An alternative is to replace the activities $\alpha_s$ with firing probabilities $p_s=\alpha_s/\lambda$ with $\lambda=\sum_q\alpha_q$ the system activity:
\begin{eqnarray}
\Delta_{i}(r\rightsquigarrow s)=\begin{cases}
p_s(i+1)-p_s(i) &\text{ if event $i$ is due to rule $r$}\\
0 &\text{ otherwise}
\end{cases}\label{eq:prob}
\end{eqnarray}
 We aggregate these contributions during the interval $[t,t+\tau]$ and divide by the number $\#r$ of events that were due to $r$ to obtain the average influence of $r$ on $s$ during that interval:
\begin{equation}
\langle\Delta_t(r\rightsquigarrow s)\rangle_{\tau}=\dfrac{1}{\#r}\sum_i \Delta_{i}(r\rightsquigarrow s) \text{ with } i\in \{j\,|\, \text{time}(j)\in[t,t+\tau]\}, \label{eq:totalinfluence3}
\end{equation}
where $\text{time}(i)$ is the time stamp of the $i$th event. We shall refer to the matrix or graph defined by (\ref{eq:totalinfluence3}) and using (\ref{eq:relative}) as the activity-DIN, or aDIN, and to (\ref{eq:totalinfluence3}) using (\ref{eq:prob}) as the probability-DIN, or pDIN.

An edge in the aDIN is easy to interpret: if at time $t$ the firing of $r$ produces patterns to which $s$ is applicable, the a-influence is positive; if it alters (and thus removes) patterns to which $s$ could have applied, it is negative. The interpretation of the pDIN is more subtle, since an $r$-event may affect the firing probability of $s$ not only by producing or destroying matchings of the pattern $\lhs_s$, but indirectly by affecting the system activity $\lambda=\sum_q\alpha_q$ through influencing the activities of rules other than $s$.

The definition (\ref{eq:totalinfluence3}) is, informally, reminiscent of a derivative $d\alpha_s/dt$ in a continuum setting, if only we could take the limit $\tau\to 0$. In general, one cannot write a system of differential equations for the evolution of the rule activities in terms of only the $\lhs$ and $\rhs$ patterns of the rules. One might therefore think of the DIN as a purely observational version of the elusive formal system of interdependencies between rule activities. We believe however that the DIN is useful in an ``empirical'' approach towards understanding a rule-based model.

\subsection{The \textit{DIN-Viz} Software Application}
\label{dinviz}

\textit{DIN-Viz} features two main view panels: the Network Panel, which presents an interactive dynamic network of rules (nodes) and the amount of influence between the rules (edges); and the Data Panel, which presents detailed information about selected rules. \autoref{fig:teaser} shows an overview image of \textit{DIN-Viz}, with the Network Panel on the left and the Data Panel on the right.

In the Network Panel, the DIN is presented as a node-link influence diagram, leveraging a force-directed layout~\cite{dwyer2009scalable} to position related nodes and clusters near to each other. The nodes and links directly represent the rules and the influence of each of the rules comprising the DIN, and the visualization enables the topological analysis of the \textit{KaSim} simulation data at the time step scale. By mapping the rule-to-rule influence to the link strength, highly mutually influential rules are closely grouped together visually. This provides insight to the systems biologist as to which rules are likely to be indicated in the formation of biological pathways. In order to further facilitate the analysis of these highly related rules, we perform a clustering operation on the network based on the influence between rules and generate groups with a high inner influence, based on a user-selected threshold.

In the Data Panel, we provide additional views that provide a temporal context for the graph representation. The upper time-based chart is the phenotype line graph, which charts \textit{KaSim} ``observables'' over the course of the simulation. The lower time-based graphs chart the incoming and outgoing influence of user-selected nodes.

\subsubsection{Network Panel}

To convey the density of information necessary for the accurate analysis of the data, we encode both data that is produced directly by the simulation and data derived from that simulation. The derived data include rule clusters, averaged influence values (when clustering globally or within a time window), and interpolated values during animation. 

To visualize the system and its data, we use a force-directed layout of a node-link diagram, where nodes encode the rules, and links encode the influence between two rules. The absolute value of influence between rules, whether defined by activity or probability (see \autoref{ss:Define}), determines the attractive link forces in the network. \textit{DIN-Viz} maps the number of hits of a rule to the node size, and the influence of one rule on another through the link width. We indicate the directionality of a link and the sign of the influence by using two directional gradients: yellow-to-red for negative, and yellow-to-green for positive. Alternatively, a colorblind safe mode is available that replaces the yellow-to-green positive colormap with white-to-blue.

Details-on-demand for the links show the source and target, as well as the exact influence value, and for the rules show the rule name, self-influence, and the top incoming and outgoing influences to the rule. The user may also choose to make visible the names of the rules as labels, either for all rules, or for interactively selected rules. 


By default, we use curved lines in order to represent edges. We rely on clustering for immediate spatial analysis and data-on-demand for more in-depth quantitative information, introducing curvature only as a minimal intervention to help differentiate multiple edges sharing source and target nodes. The use of curves does not appear to detract from the intensity gradient used to encode direction~\cite{holten2009user}. However, the literature on graph perception indicates that curved edges can hinder readability~\cite{xu2012user}, so we provide an option for users to toggle between curved or straight edges.

\begin{figure}[!t]
\begin{center}
\includegraphics[width=\columnwidth]{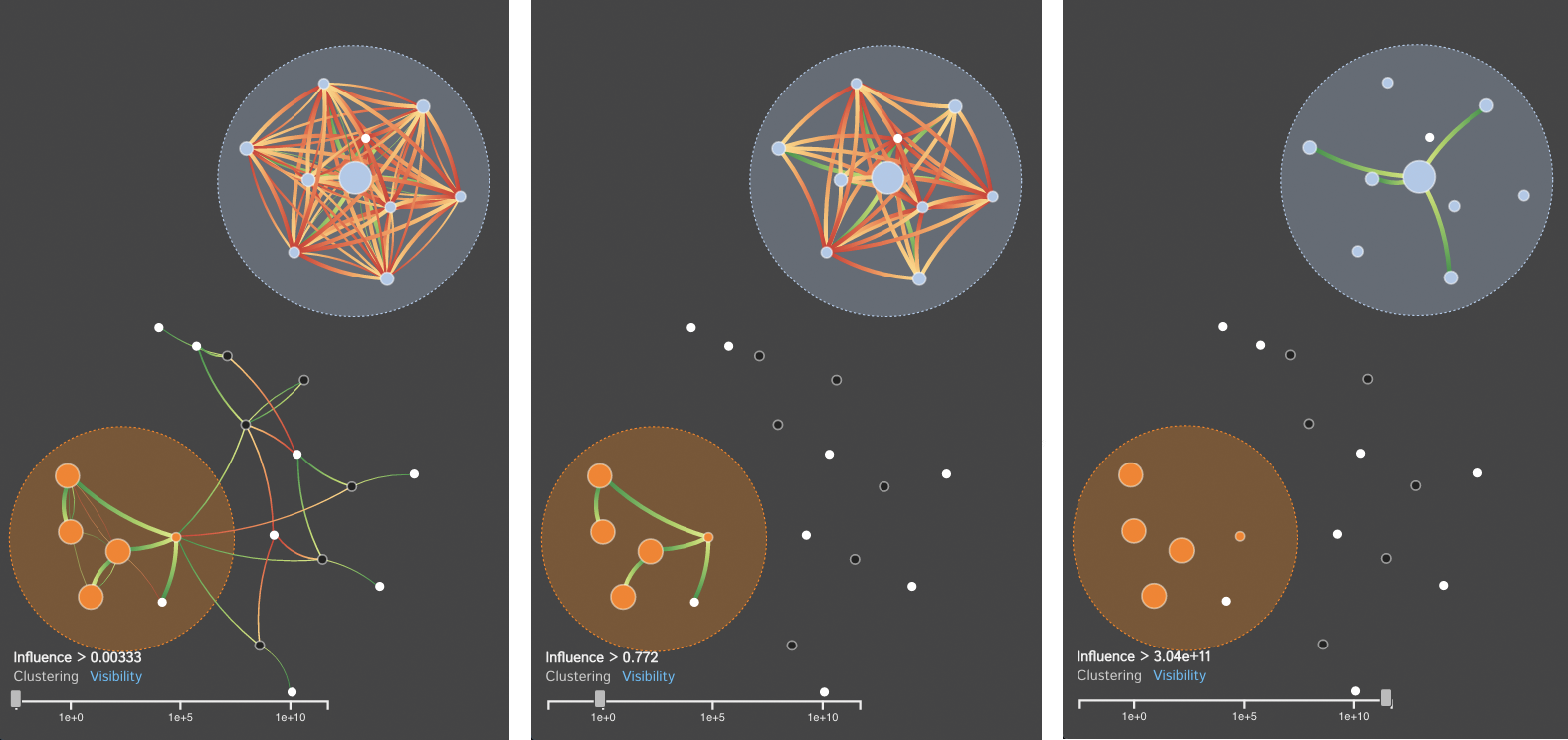}
\caption[Vis]{Links may be interactively filtered on demand using the visibility slider (bottom-left of each image). With all links visible (left), the light blue cluster is very highly interconnected. Increasing the visibility threshold (middle) causes the free nodes to have no visible links, showing that they do not have a strong relationship any nodes at this time step. At the highest visibility threshold (right), only 4 links remain, all connected to the node in the center of the light blue cluster. These important relationships are difficult to see when more links are visible (left, middle).}
\label{ThresholdVisibility1}
\end{center}
\vspace{-.5cm}
\end{figure}

\paragraph{Rule Clustering} We derive cluster assignments for the rules based on their influence upon one another. To form these clusters, we extract all links from the network. If a link is above the user-determined clustering threshold, its source and target rules are considered to be clustered together. If nether is currently associated with a cluster, they are placed together in a newly created cluster. If one is in a cluster and the other is not, then the unclustered node joins the cluster. Lastly, if both nodes currently belong to different clusters, these two clusters are joined together.

Onto these derived clusters we map a categorical color scheme to distinguish between clusters. We double encode the cluster assignment through the node color as well as underlaying clustered nodes with a bounding circle of the cluster color. Double encoding is used to emphasize the relationship between nodes; in our iterative design process, we discovered that simply changing the color of nodes did not sufficiently highlight them, especially inside of dense clusters.

Clustering is helpful for grouping rules together and aids the user in understanding how rules operate together. However, these simulations can have influence values spanning orders of magnitude. To handle diverse simulations, the user can fine-tune the clustering threshold based on the nature of the system, as well as the interactions they would like to highlight. In medium-sized or smaller networks (with less than 500 nodes), the clustering computation can be performed at real-time rates on a consumer laptop. The responsiveness of the clustering method provides strong interactivity with the visual analytics tool.

The initial clustering mode is per time step, where the clusters will be computed using links from the current time step. This operation yields very precise groupings of rules based on their current states. The clusters can be very volatile for certain systems, and a node could potentially switch between clusters at every time step. To provide an alternative to this, we include two additional clustering modes: global and windowed. These modes can smooth the clustering operation when moving backward or forward in time, and can help the user to highlight trends across time.

\begin{figure}[!t]
\begin{center}
\includegraphics[width=\columnwidth]{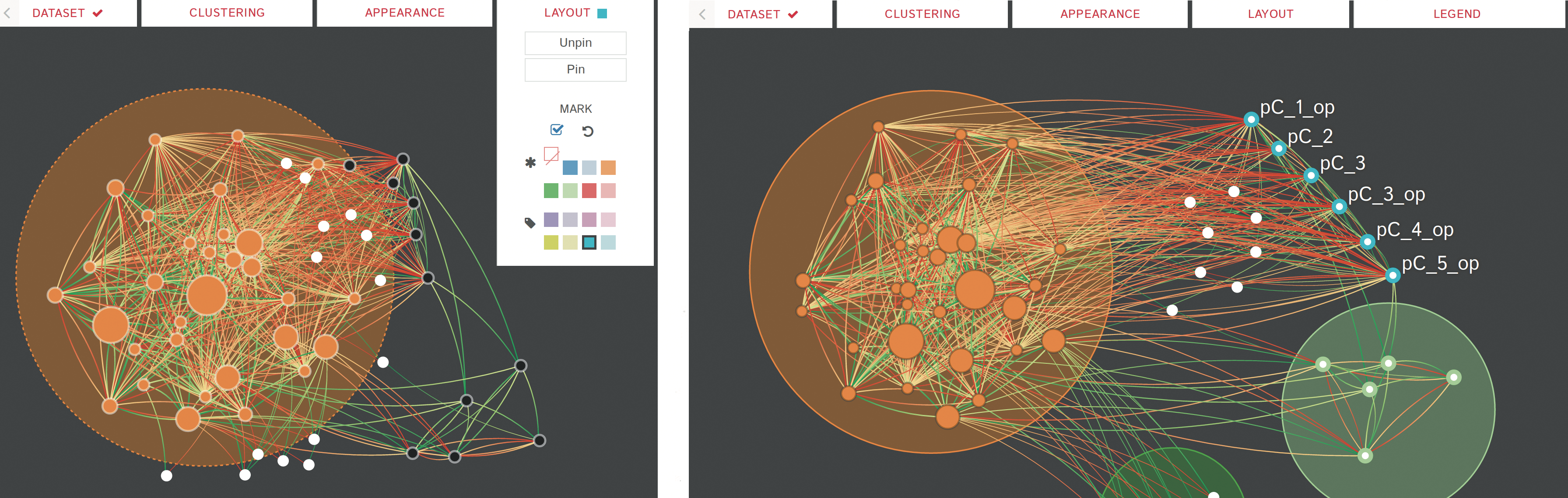}
\caption[MarkAndPin1]{Influence networks can become very complicated (left). Through pinning and marking, the user may apply custom spatial and categorical groupings that serve to unclutter as well as annotate the rules. After reorganization (right), rules pC\_* are marked and labeled, as positioned near to one another. The white nodes below and to the right of the orange cluster (left) have been pinned outside of the orange cluster with two new marked clusters being created, green and light-green (right). A pop-up menu provides options for the pinning and marking features.}
\label{MarkAndPin1}
\end{center}
\vspace{-.1cm}
\end{figure}

Global clustering averages the influence that all nodes have on one another over all time steps and transforms these average influence values into the links used to cluster nodes. Globally clustering the rules groups them by their overall behavior, not just at one specific instant. A downside to this approach would be a link which fluctuates perfectly from a value $x$ to $-x$ over the course of the simulation, resulting in an average influence of $0$. However, the clusters fail to capture the more granular behavior may when attempting to group rules globally. As a middle ground, we implement a time-window clustering that smooths volatility while preserving detail. The user can specify the number of time steps that the clustering will use. For example, if the user selects a window of 10, the influences are averaged over 21 time steps $[t-10, t+10]$, including the current time step. This method preserves the local behaviors while smoothing the cluster assignments over time.

By changing the clustering window, the user can analyze the data at different time scales. \autoref{ThresholdCluster1} shows a view of the DIN at the same time step, but using three different clustering thresholds. As the user increases the clustering threshold value (using the slider at the bottom-left of the Network Panel), fewer and fewer clusters are created. In this figure, the user has selected a time window of 5, so that clusters are determined through averaging influences over the 5 previous and 5 subsequent time steps.

\paragraph{Interactive Filtering} We also allow for the interactive filtering of links by their influence value to help handle larger datasets. By removing the edges below a user defined influence value, the graph is less visually cluttered, and the edges with higher influence values become prominent, whereas they could otherwise be obscured. \autoref{ThresholdVisibility1} shows an example where a user increases the visibility threshold. As the user increases the visibility threshold value (again, using the slider at the bottom-left of the Network Panel), more and more edges are removed from the visualization. For example, in the rightmost frame, the four positive influences are clearly seen within the larger gray cluster once the negative edges with less influence are  removed.

\paragraph{Marking and Pinning} While force-directed layouts help to mitigate visual clutter in node-link diagrams, dense networks can still be difficult to make sense of--- an issue that is exacerbated when representing large datasets. \textit{DIN-Viz} includes an option whereby the user can manually creates a layout of nodes or entire clusters through relocating and ``pinning'' them to specified locations in the Network Panel. This reorganization reduces clutter, but also helps users to distribute the rules and clusters in a way that is cohesive with their thought process during exploratory analysis. When pinned, the spatial positioning and grouping of selected rules and clusters is preserved over the course of the entire simulation, overriding the normal layout behavior.

While the mathematically defined clusters help capture groups of rules with a strong influence on one another, there may be rules which are related through their behavior but lack a strong influence with one another. To solve this problem, we implement a ``painting,'' or marking, interaction. The user can mark a set of nodes with the same color to give them a categorical grouping, inserting them into an existing or newly created painted cluster. Alternatively, nodes can be painted and labeled, grouping these rules visually, but only with color, not placed within a cluster. These categorical groupings, similar to the spatial groupings achieved through pinning, aid in the logical organization of rules during analysis by the user.

Through pinning and marking actions, users are aided in the exploration of the dataset, leading to the formation and evaluation of hypotheses. Additionally, pinning and marking, along with displaying the labels of rules, serve as tools to annotate the data and prepare it for presentation. Grouping rules both spatially and categorically can help to emphasize specific interactions and biological pathways. \autoref{MarkAndPin1} shows an example of how the use of marking and pinning can be used to effectively reorganize a dense influence network to highlight specific sets of rules. The operations of labeling, cluster windowing, cluster thresholding, visibility thresholding, pinning, and marking, can all be used in combination to articulate components of the network that are relevant to a particular analysis session.

\begin{figure}[!t]
\begin{center}
\includegraphics[width=\columnwidth]{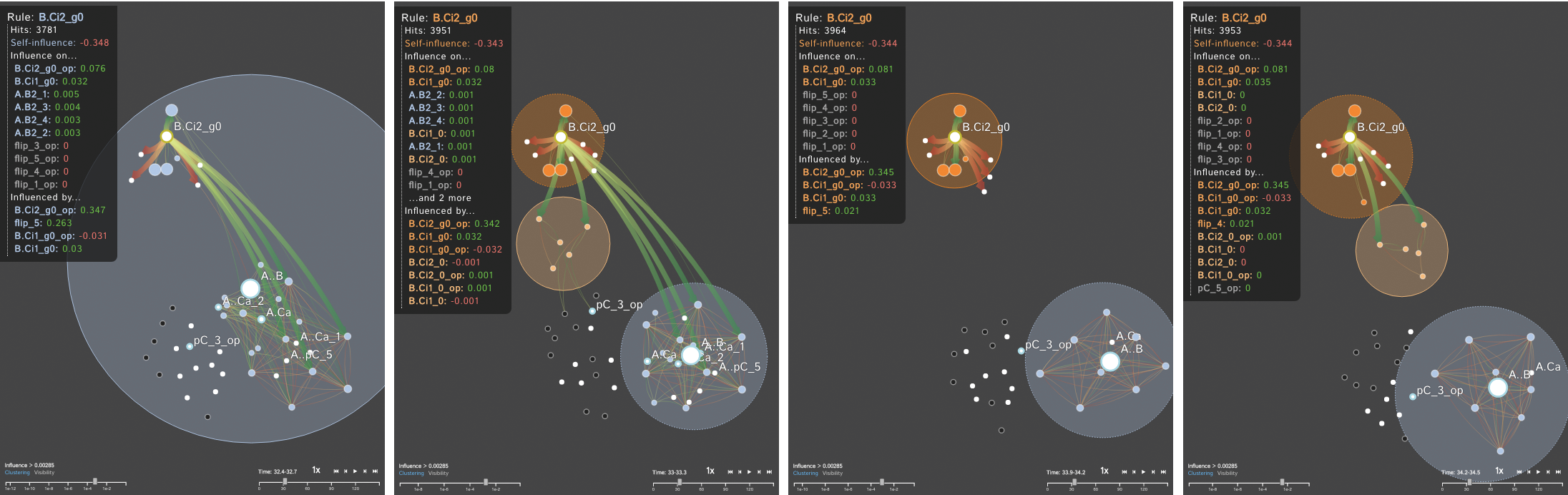}
\caption[timeline]{These four screen captures show the changes in rule influence as the user steps through the simulation in the Network Panel. Here, the clustering of rules changes as each rule's influence increases or decreases, and the incoming and outgoing influences of a rule chosen by the user (here the rule\textit{B.Ci2\_g0} is selected) are indicated both by the highlighted edges and the pop-up panel in the upper-left corner of the visualization. Additionally, a dashed indicator marks the corresponding time in the line graphs (See~\autoref{fig:teaser}), linking network and time-based contexts together. An additional controller below the graphs allows the user to zoom into a smaller time window for a more detailed view.}
\label{timeline}
\end{center}
\vspace{-.1cm}
\end{figure}

\paragraph{Animation} To represent the dynamism of the system, animation is used to update the influence between the rules, which in turn updates the edge weights, node sizes, and cluster definitions. A time slider controls the current time step, enabling the user to move through time or to jump to a particular time step. We also include standard playback controls that animate the simulation so that the user can observe changes in the DIN over time. The user can play and pause the animation, as well as speed up or slow down the animation. The \textit{KaSim} simulation data provides ``snapshots'' at particular time intervals. Playback across longer intervals could cause confusing jumps where the network suddenly leaps to a completely new configuration. While this may be important to visualize (and encourage the systems biologist to re-run the simulation at a greater temporal resolution), even less dramatic shifts can obscure the dynamics of the system. To improve the temporal coherence as we animate between frames, we provide an option where we perform a linear interpolation of influence values between two sequential time steps~\cite{benard2011state,villegas2015hvei}. \autoref{timeline} demonstrates the evolution of a DIN over time, depicting how the clusters of rules emerge and disappear as the rules' influence increases and decreases.

\subsubsection{Data Panel}

Although the use of animation helps to facilitate insight into the dynamics of the system, it can make it difficult to keep track of detailed differences between time-steps, hindering comparative analysis and increasing cognitive load~\cite{beck2017taxonomyCGF,tversky2002animation}. To address this problem, we supplement the Network Panel with the Data Panel. 

The Data Panel provides the explicit temporal context missing from the Network Panel, enabling a thorough multi-level analysis of the data. The Data Panel consists of 3 line charts, graphing information dependent on time. The Phenotype Chart graphs observables--- populations of particular biological elements--- from the simulation over time. This view assists in the analysis of global trends in the simulation for the entire system. The additional Rule Influence Charts display the incoming and outgoing influences of a user-selected node. The Rule Influence Charts show global trends in the simulation for a single rule, rather than the entire system. Self-influences are denoted by a red line, while influences to and from other rules are denoted by gray lines. Hovering over these lines shows the corresponding time step and influence values while highlighting the line itself. By using the three charts in the Data Panel, system trends over time may be analyzed at both a global and local level simultaneously. Located at the bottom of the Data Panel is the time selection slider, which enables the user to zoom the time axes of the Phenotype and Rule Influence Charts. The right side of \autoref{fig:teaser} shows the Data Panel.

\section{Case Study} \label{UseCaseSection}

To evaluate the DIN technique, we present a detailed use case in which a systems biologist examines the dynamics of a complex protein-protein interaction network. The basic usage practice consists in: (i) Visualizing the dynamics of the influence network and identifying clusters of rules,  determined by mutual strength of influence, regardless of whether influence has a positive or negative sign; (ii) Identifying temporal sequences of activation and inhibition between rules or rule clusters; and (iii) Using these temporal ``story lines'' to connect the mechanisms laid down in the rules with a specific macroscopic behavior of the system. For our use case, this practice occurred in iterative steps, as, for instance, an identification of an interesting sequence (ii) could lead to the need to re-conceptualize which rules were most relevant (i), or a hypothesis about the meaning of a story line (iii) could require a change to the pinning or thresholding of rule clusters in order to reveal potentially meaningful dynamics (ii).


\subsection{The KaiABC Oscillator}
\label{ss:KaiABCoscillator}
Many organisms control physiological processes in a pattern that anticipates daily changes in light levels. Such circadian rhythms are maintained by molecular networks that exhibit autonomous oscillations that directly involve or are coupled to gene regulation. A widely studied system is the circadian oscillator of the cyanobacterium \emph{Synechococcus elongatus}~\cite{nakajima2005reconstitution}. This system is remarkable because its key components are just three proteins--- KaiA, KaiB, and KaiC--- which have been isolated and reconstituted in the test tube, where they maintain the oscillating behavior under consumption of chemical energy (ATP)~\cite{van2007allosteric}. Though this rule-based system is relatively small, it exhibits many of the complexities found in much larger models.


\begin{enumerate}[noitemsep,leftmargin=*]
\item
The protein KaiC is a complex of six identical units each of which can be reversibly phosphorylated (marked with a phosphate group) independently of the others, with dephosphorylation (unmarking) occurring more readily than phosphorylation (marking). The hexamer as a whole (henceforth simply KaiC) has therefore $6$ sites and $2^6=64$ possible phosphorylation states, but we distinguish only $7$ overall phosphorylation levels ($p$-levels) $p\in\{0,1,\ldots,6\}$. 

\item
 KaiC is assumed to switch between two conformational states, termed ``active" (A) and ``inactive" (I). Importantly, the probability (i.e.\@ the rate constant) of a conformational flip from A to I increases with the $p$-level of KaiC.

\item
\label{three}
KaiA binds KaiC in the A-form, but the interaction strength (more specifically: the average time that KaiA stays bound to KaiC) decreases rapidly with increasing $p$-level.  When bound, KaiA prevents conformational flipping and promotes the phosphorylation of KaiC. In the model, this is done by upping the phosphorylation rate and preventing dephosphorylation. In sum, KaiA sticks to KaiC in the A-form at low $p$-levels, cranks up the $p$-level, and by doing so kicks itself off KaiC. When KaiA unbinds from KaiC at high $p$-levels, KaiC flips back into the I-form, which KaiA cannot bind. At this point the dephosphorylation tendency takes over, leading to a decrease in the $p$-level. 
\end{enumerate}

At low numbers of KaiA, (1)-(3) induce individual KaiC molecules to move up and down $p$-levels rather randomly and without coordination among each other. As a result, the population of KaiC molecules settles on a steady level of average overall phosphorylation. 

\begin{enumerate}[noitemsep,leftmargin=*]
\setcounter{enumi}{3}
\item
\label{four}
Coordination between KaiC proteins is achieved by a third agent, KaiB, which reversibly binds KaiC in the I-form. When bound, KaiB locks KaiC in the I-form, thereby allowing dephosphorylation to proceed. Importantly, bound KaiB also strongly (but reversibly) binds KaiA. The probability that KaiA binds KaiB, when the latter is bound to KaiC, is maximal at intermediate $p$-levels (but the time it stays bound is independent of $p$-level). While bound in this way, KaiA can no longer interact with any KaiC. Hence, KaiB promotes dephosphorylation of the KaiC it is bound to and holds back phosphorylation of other KaiC molecules by sequestering KaiA. These interactions statistically synchronize the phosphorylation and dephosphorylation cycles on individual KaiC molecules, resulting in collective $p$-level oscillations of the KaiC population. 
\end{enumerate}


When translating these mechanistic descriptions (some of which are empirical findings and some of which are hypotheses~\cite{van2007allosteric}) into \textit{Kappa}, we obtain $57$ rules, whose dynamical action is simulated using \textit{KaSim}. If the same model were based on a description in terms of molecular species instead of patterns, it would require $707$ species and $13090$ reactions. In the present case, KaiC is represented as an agent with $6$ phosphorylatable sites, because \textit{Kappa} is (deliberately) not designed to express the idea of equivalent sites or the notion of a phosphorylation ``level.'' However, we can choose to automatically detect such equivalences (and hence implicitly the notion of a phosphorylation level). In so doing, the number of species shrinks to $59$ and the number of reactions they participate in turns out to be exactly $1000$. Even in relatively simple cases as this one, the rule-based approach provides a hugely compressed description that is lossless with regard to dynamics.


\subsection{Using DINs to Understand the KaiABC Clock}
\label{ss:understandKaiABC}

\begin{figure}[!t]
\begin{center}
\includegraphics[width=\columnwidth]{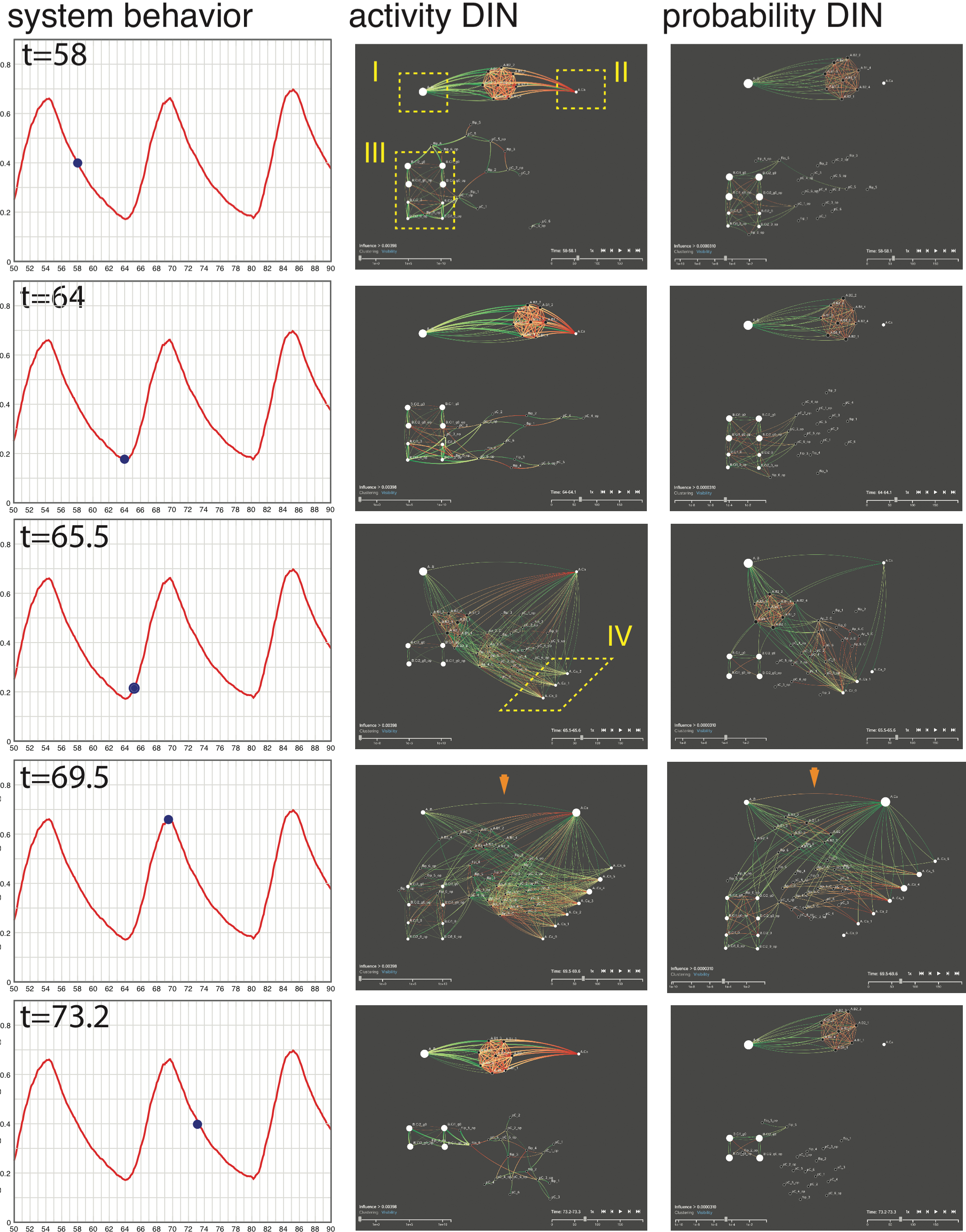}
\caption[fig:snap]{This figure shows a series of snapshots taken during an analysis session of the KaiABC clock, discussed in \autoref{ss:understandKaiABC}. The left column shows the current time step and the global phosphorylation in the system (the $y$-axis) across the simulation session, with a point indicating the particular time step (the $t$-axis). The middle column presents snapshots of the DIN where influence is defined using the relative change in activity (the \textit{aDIN}). The right column presents snapshots of the DIN where influence is defined based on the firing probabilities of rules (the \textit{pDIN}).  
}
\label{fig:snap}
\end{center}
\vspace{-.5cm}
\end{figure}

After spending an initial period of time becoming familiar with the overall dynamics of the system, we then begin a more focused visual analysis to identify important features, or ``story lines,'' of the KaiABC clock that are made apparent using \textit{DIN-Viz}. \autoref{fig:snap} depicts a series of time steps from this analysis session and is referred to throughout this section.

We proceed first by turning off visual clustering, working only with the default structure imparted by the force-directed layout. We also turn off all edges, leaving only unmarked nodes. Running the DIN animation in this terse format allows quick identification of which nodes (rules) are potential drivers by paying attention to (a) the evolution of node size, which reflects the number of firings in the corresponding time interval (displayed on the lower-right), and (b) identifying nodes that move abruptly or rapidly, as such nodes appear to signal changes in the influence pattern. A case in point is the node {\tt A.Ca} in the timeframes $[64.6-64.7]$ to $[64.7-64.8]$. Around $t=64.6$, {\tt A.Ca} had no firings but received a high positive influence from rule {\tt A..B}. At $t=64.7$, {\tt A.Ca} starts firing and suddenly connects with a host of other nodes, as can be seen when turning on the edge display. Using the full graph display is, of course, useful too, in particular when keeping an eye on the trajectory of the systemic property shown in the right panel of the browser. It is also informative to turn on and off all inhibitory edges (red) or all activating edges (green).

The next step consists in marking and pinning nodes that appeared salient in the first pass. It  proves useful to rerun the animation before pinning in order to determine any cluster structure among the salient nodes, which would suggest a division of the screen area into regions dedicated to these clusters. It is important to keep in mind that cluster structure is influence-based and therefore likely transient. The purpose of pinning is to fix a ``coordinate'' system of salient nodes against which to assess the tug of war of influence over the remaining mobile nodes. If the pinned nodes are indeed salient, they should engage in distinct patterns of activity that organize the mobile nodes. Tuning the visual clustering (by setting the cluster threshold) can be useful to determine cohesive structure among the mobile nodes. Note that here pinned nodes are excluded from such visual clustering.

\autoref{fig:snap} exhibits our layout for the KaiABC clock. In this model several rules have a natural family structure. For example the {\tt A..Ca\_x} rules represent the dissociation of $\ce{A}$ from an active $\ce{C}$ at a rate constant that depends on the $p$-level of the $\ce{C}$-agent. These rules are all expressing the same basic transformation (unbinding of $\ce{A}$ from $\ce{C}$) but their refinement according to $p$-level is a critical part of the differential affinity mechanism~\cite{van2007allosteric}. Often, but not always, do such rules cluster naturally in the DIN. A modeler should therefore make use of any information he or she has about the nature of the rules to group them using \textit{DIN-Viz}.

Once parts of the DIN are fixed, the state of the DIN needs to be put into correspondence with the macroscopic behavior of the system. For this purpose we supply a file with a time series of the property of interest (i.e., a Phenotype Chart). In our case, the systemic property is the overall fractional phosphorylation $Y$:
\begin{equation}
Y=\dfrac{\sum_i^n i |\ce{C(p=i)}|}{n |\ce{C()}|}
\end{equation}
where $|\ce{C(p=i)}|$ is the number of $\ce{C}$-agents with $i$ phosphorylated sites, $|\ce{C()}|$ is the total number of $\ce{C}$-agents, and $n$ the number of phosphorylatable sites---in our case $n=6$. The quantity $Y$ can be viewed as the probability that any of these sites is phosphorylated. $Y=1$ means that all $\ce{C}$-agents are fully phosphorylated, and $Y=0$ means that none has even a single phosphorylated site. The left-hand column of \autoref{fig:snap} shows a few oscillations of $Y(t)$ with a dot marking the position in the cycle corresponding to the aDIN and pDIN snapshots on the right. 


The top row ($t=58$) shows 3 of the 4 anchor clusters of nodes we select by observing the animation as described above. While the initial confidence in these selections is supported by detailed knowledge of the model, we believe that someone without knowledge of the model would have discovered the same selection. First, the selected nodes are conspicuous in the temporal evolution of their firing rate and, second, the mobile nodes are seen to be meaningfully organized by the influence behavior of these anchors. The latter is an after-the-fact property that may require some iteration, which this interactive approach to DIN visualization is designed to facilitate.
Anchors I and II (\autoref{fig:snap}, middle) are the single rules {\tt A..B} and {\tt A.Ca}, respectively. Rule {\tt A.Ca} is central to the processes described in item \ref{three} of \autoref{ss:KaiABCoscillator}, while rule {\tt A..B} and the anchor cluster III are central to those described in item \ref{four}. In the down-leg of the system cycle, many of the $\ce{C}$-agents are in a state where they are being dephosphorylated, as they are locked into the inactive conformation by the binding of $\ce{B}$ (anchor III). 

Interacting with anchor III is a cloud of mobile nodes containing rules named {\tt pc\_x\_op}, which are dephosphorylation rules. (The {\tt x} denotes the site on which the rule operates.) In addition, the cloud contains {\tt flip\_x} rules, which flip a $\ce{C}$-agent from the active to the inactive state, promoting dephosphorylation. (The {\tt x} in the rule name denotes the $p$-level of the agent, not a specific site.) Finally, the cloud also contains {\tt pC\_x} rules, which are phosphorylation rules.  The animation makes sense of all this: One can see a ``catching'' motion where sites on $\ce{C}$-agents are dephosphorylated, ultimately leading to a $p$-level of zero, turning on the {\tt B.Cix\_0\_op} nodes, which effectively cause agents of type $\ce{B}$ to dissociate from agents of type $\ce{C}$, initializing the phosphorylation phase. At the same time, {\tt pC\_x} rules fire and activate {\tt flip\_x} rules, indicating that some $\ce{C}$ are still in the previous up-leg of the cycle where they are reaching the peak and get switched into the inactive state by binding to $\ce{B}$ (rules {\tt B.Cix\_g0}, which fire at higher $p$-levels). This analysis of the ``catching'' motion upon animation (at any edge visibility threshold) simply indicates the flow along the down-leg of the cycle: A majority of the $\ce{C}$ population has entered the downward leg and is dephosphorylating, thus promoting eventual detachment from $\ce{B}$, which locks the inactive state. However, another fraction of the population is lagging behind, still entering phosphorylation peak territory. The microscopic picture around anchor III enables these conclusions, although the systems level data simply indicate an overall declining $p$-level of the population.  

According to the DIN, the processes that dominate at the midpoint of the down-leg have not changed much their interrelations when the system-level trajectory has reached the valley floor at $t=64$. Consider now what is going on, in parallel, between anchors I and II at $t=58$ or $t=64$. As indicated in item \ref{four} of \autoref{ss:KaiABCoscillator}, the phosphorylation of active $\ce{C}$-agents is promoted by their binding to agents of type $\ce{A}$, rule {\tt A.Ca}, which is our anchor II. However, agents of type $\ce{A}$ are sequestered by agents of type $\ce{B}$ when the latter are bound to highly phosphorylated $\ce{C}$. The point of this is to allow dephosphorylation to proceed by silencing the $\ce{A}$-agents (which promote phosphorylation). But once $\ce{C}$-agents have reached maximal dephosphorylation levels, they should enter the up-leg, for which the $\ce{A}$-agents need to be awake. This is accomplished by releasing (unbinding) the $\ce{A}$-agents through rule {\tt A..B}, anchor I. The hairball in the middle is a cluster of mobile nodes that are about the binding of $\ce{A}$ to $\ce{B}$, the opposite of what \mbox{anchor~I} accomplishes. By definition of influence (\autoref{ss:Define}), the firing of {\tt A..B} has a positive effect on the binding rules {\tt A.Bx} because it makes $\ce{A}$ available. In the downward leg of the cycle, {\tt A..B} fires much more frequently (the large disk) than {\tt A.Bx}, but the latter is still firing with appreciable frequency in this snapshot---presumably because of the laggards that are now reaching the peak and ready to bind $\ce{B}$, lock in the inactive state and sequester $\ce{A}$. 

Rules {\tt A.Ca} and {\tt A.Bx} compete for free $\ce{A}$, rendered by the red arrow from {\tt A.Bx} to {\tt A.Ca}. Yet, at $t=64$ the negative influence is not reciprocated, because {\tt A.Ca} doesn't fire at all, and this despite the availability of activated $\ce{C}$ (or {\tt A.Ca} could not be influenced). When stepping through the DIN at $t=65.5$ and beyond, one notices that {\tt A.Ca} never reciprocates the negative influence, even when it begins to fire. How can that be, if, from a static analysis point of view, {\tt A.Ca} removes a free $\ce{A}$ from the system that could have also participated in {\tt A.Bx}? The reason is that the binding rate constant of {\tt A.Bx} is so high (both in the real system and the model) that once a $\ce{B}$ binds to a $\ce{C}$, creating the condition for an $\ce{A}$ to bind that $\ce{B}$, the event happens very quickly. There rarely is just a $\ce{B}$ bound to a $\ce{C}$ in the system. In other words: the rule {\tt A.Bx} fires to immediately consume any pattern instance to which it is applicable. This creates the seemingly paradoxical situation of a rule that fires often and has almost always activity zero. A rule with activity zero cannot be negatively influenced; it can only be influenced positively, on the supply side, i.e.\@ by {\tt A..B}. The DIN shows that, at this point in the cycle, whenever {\tt A..B} feeds the system with free $\ce{A}$ (green arrows oriented towards {\tt A.Bx}), {\tt A.Bx} makes sure nothing of that reaches {\tt A.Ca} (red arrows from {\tt A.Bx} to {\tt A.Ca}). 

The DIN panel at $t=58$ shows two parallel mechanisms at work in the downward leg of the Kai oscillator. (i) The binding of $\ce{B}$ to $\ce{C}$ locks $\ce{C}$ into its inactive state and allows dephosphorylation to proceed more effectively. This mechanism is at work in the lower half of the DIN panels. (ii) The high affinity of $\ce{A}$ for $\ce{B}$ bound to inactive $\ce{C}$ puts the breaks on the front-runners, i.e.\@ those $\ce{C}$-agents that have already reached the valley floor and switched into the active state, ready for the upward (phosphorylation) leg of the cycle. Those agents need rule {\tt A.Ca} to fire and bind them to $\ce{A}$, locking their active state and promoting phosphorylation. But despite these very $\ce{C}$-agents having released $\ce{B}$, which in turn releases $\ce{A}$ (high firing rate of  {\tt A..B}), $\ce{A}$ gets immediately captured by $\ce{B}$-agents bound to laggard $\ce{C}$-agents that have not yet completed their dephosphorylation leg. This holds back the front-runners ready from initiating the upward leg while ensuring that the laggards complete their downward leg. The upper part of the DIN reproduces precisely the statistical synchronization mechanism proposed by van Zon et al.~\cite{van2007allosteric}, but seen through the lens of a rule-based model. The fact that group IV is already active at $t = 65.5$, a time when the system-wide $p$-level just passed its minimum, indicates how spread out the distribution of individual $p$-levels must be, as some are evidently already nearing the top. The basic structure of the DIN then persists throughout the upward phase and beyond, before it breaks down again at $t = 73.2$ and the previous organization reappears.


In sum, the use case illustrates a system in which KaiC proteins go through a phosphorylation cycle. They are incrementally phosphorylated at 6 positions in the molecule when they switch to a dephosphorylation process that removes the phosphates. Yet, even if each protein goes through such a cycle, this would not lead to synchronization across the whole population of proteins. In the model, a series of interactions causes a large fraction of individuals to synchronize and this causes oscillations at the level of the whole population. Using the DIN, the systems biologist can begin to investigate subtleties of the  dynamics of these interactions in order to better understand this population-level synchronization.

\section{User Feedback} \label{DiscussionSection}

While the detailed case study presented above provides a real-world example of how a biologist can analyze a rule-based model using DINs, 
we also gathered initial feedback regarding the \textit{DIN-Viz} tool, both through unstructured interviews and written responses to questions about how \textit{DIN-Viz} might be used to support relevant analysis tasks. Feedback was provided by professors and postdoctoral researchers in systems biology or molecular biology from Harvard Medical School and University of Arizona. Each of the six researchers who provided feedback are actively engaged in investigating protein-protein interaction networks related to cancer biology and interested in using bioinformatics tools for rule-based modeling of biological networks. 

User response was, for the most part, overwhelmingly positive. For example, one user was initially skeptical since the KaiC use case merely ``enables recovery of what we already know.'' However, after exploring the model in more detail, the user expressed more interest, telling us he was impressed by the ability to reason about macro and micro level behaviors simultaneously: ``What I find spectacular is this juxtaposition of the crazy moving hairball on the left, which is the microscopic picture, and the quiet calm coordinated oscillations at the systems level on the right. That is really something emotional to see the two together.'' One of our systems biology collaborators told us that they were very intrigued by being able to look alternately at the aDIN and pDIN versions of the simulation: ``I think I now see the subtle footprint of a critical synchronization mechanism at work in the Kai clock. Seeing is an active act, which is sometimes in the way; the DIN visualization switches ``seeing'' into a receptive mode where you can get rid of the bias that comes from the active version of seeing.'' Another user was excited about the potential pedagogical utility of \textit{DIN-Viz}: ``The coloring and pinning of the nodes is terrific! A very useful way of interacting with the tool. I could learn a great deal about this system by working with this visualization. I think it’s going to be very useful in the classroom.'' However, we were also cautioned that our use case provided only an initial confirmation of the success of our approach: ``You need to make a model of a system that's not yet understood and show how the visualization enables a bona fide explanation.'' 

While encouraged by this feedback, in future work we will more rigorously evaluate the use of DINs to help make sense of a wide range of biological processes, both in scientific and educational contexts. In general, there is a lack of evaluation about the effectiveness of rule-based approaches, and \textit{DIN-Viz} could be used to help understand when it is most appropriate to use them to make sense of complex systems.

\section{Conclusions} \label{ConclusionSection}

In this paper, we introduced the Dynamic Influence Network, a novel visual analytics technique for investigating the dynamics of how rules influence other rules in rule-based models of complex systems. We demonstrated that our web-based application, \textit{DIN-Viz}, supports a range of visual analysis tasks for making sense of dynamic protein-protein interaction networks.

While our approach has shown to be useful for the manual exploration of these types of rule-based modeling systems, our initial evaluation leads us to believe that some of these manual explorations could be automated. That is, just as a researcher observes that particular rules fire at similar rates, an automated system might highlight a range of interesting temporal patterns, allowing the user to investigate whether or not these patterns are scientifically meaningful.


The goal of the DIN is precisely to focus on rules rather than individual elements or types. However, it could be useful to be able to ``drill down'' from a selected rule or set of rules to another representation, such as a contact map or biological pathway, that provided additional context for the rule dynamics. Determining how to navigate effectively between these levels of abstraction is a significant visualization challenge that we plan to investigate.





The DIN technique was designed to support the \textit{Kappa} language, and specifically to ingest \textit{KaSim} simulation data, but we expect that it should be relatively straightforward to adapt our technique to support other rule-based modeling languages. More broadly, DINs could be used to represent a wide range of dynamic systems explored in other domains, such as economics~\cite{kuhn2016rule}, ecology~\cite{filatova2013spatial}, or social network analysis~\cite{prskawetz2017role}, among others.

\acknowledgments{
This work is funded by the DARPA Big Mechanism Program under ARO contracts W911NF-14-1-0367 and W911NF-14-1-0395.
}

\bibliographystyle{abbrv-doi-hyperref-narrow}
\bibliography{DIN_References}
\end{document}